\documentclass[preprint]{aastex}    

\newcommand{\sk}{Sk~$-67{\arcdeg}111$}
\newcommand{\av}{Sk~80}
\newcommand{\iue}{\it IUE\/}
\newcommand{\hst}{\it HST\/}
\newcommand{\fuse}{\it FUSE\/}
\newcommand{\orfeus}{\it ORFEUS\/}
\newcommand{\vinf}{${v}_\infty$}
\newcommand{\kms}{km\,s$^{-1}$}
\newcommand{\teff}{$T_{\rm eff}$}
\newcommand{\ebv}{$E$(\bv)}
\newcommand{\msunpyr}{${\rm M}_\odot / {\rm yr}$}
\newcommand{\grav}{$\log\,g$}

\slugcomment{To appear in FUSE special edition of Ap.J. Letters}
\shorttitle{FUSE Observations of O7 Supergiant Winds}
\shortauthors{Fullerton et al.}
\begin{document}

\title{{\fuse} Observations of the Stellar Winds of Two O7 Supergiants in
       the Magellanic Clouds}

\author{ A.~W.~Fullerton\altaffilmark{1,2},
         P.~A.~Crowther\altaffilmark{3},
	 O.~De~Marco\altaffilmark{3},
         J.~B.~Hutchings\altaffilmark{4},
         L.~Bianchi\altaffilmark{2,5},
         K.~R.~Brownsberger\altaffilmark{6},
         D.~L.~Massa\altaffilmark{7},
	 D.~C.~Morton\altaffilmark{4},
         B.~L.~Rachford\altaffilmark{6},
         T.~P.~Snow\altaffilmark{6},
         G.~Sonneborn\altaffilmark{8},
	 J.~Tumlinson\altaffilmark{6},
    and  A.~J.~Willis\altaffilmark{3}        }

\altaffiltext{1}{Dept. of Physics \& Astronomy, 
                 University of Victoria,
                 P.O. Box 3055, 
                 Victoria, BC, V8W 3P6, 
                 Canada. { Email: \rm{awf@pha.jhu.edu} } }
\altaffiltext{2}{Dept. of Physics \& Astronomy, 
                 The Johns Hopkins University,
                 3400 N. Charles Street, 
                 Baltimore, MD 21286}
\altaffiltext{3}{Dept. of Physics \& Astronomy,
                 University College London,
                 Gower Street, London WC1E~6BT,
                 England}
\altaffiltext{4}{Herzberg Institute of Astrophysics,
		 National Research Council of Canada,
                 5071 West Saanich Road,
                 Victoria, BC V8X 4M6,
                 Canada}
\altaffiltext{5}{Osservatorio Astronomico di Torino,
                 I-10025 Pino Torinese (TO),
                 Italy}
\altaffiltext{6}{Center for Astrophysics and Space Astronomy,
                 University of Colorado at Boulder,
                 Campus Box 389,
                 Boulder, CO 80309}
\altaffiltext{7}{Raytheon ITSS,
                 NASA's Goddard Space Flight Center,
                 Code 681, Greenbelt, MD 20771}
\altaffiltext{8}{Laboratory for Astronomy and Solar Physics,
                 NASA's Goddard Space Flight Center,
                 Code 681, Greenbelt, MD 20771}

\begin{abstract}
We compare the stellar wind features in far-UV spectra of {\sk}, an
{O7 Ib(f)} star in the LMC, with {\av}, an {O7 Iaf$+$} star in the SMC.
The most striking differences are that {\av} has a substantially 
lower terminal velocity, much weaker {\ion{O}{6}} absorption, and stronger
\ion{S}{4} emission.
We have used line-blanketed, hydrodynamic, non-LTE atmospheric models 
to explore the origin of these differences.
The far-UV spectra require systematically lower stellar 
temperatures than previous determinations for O7 supergiants derived 
from plane-parallel, hydrostatic models of photospheric line profiles.
At these temperatures, the {\ion{O}{6}} in {\sk} must be due primarily
to shocks in the wind.
\end{abstract}

\keywords{stars: early-type --
          stars: winds, outflows --
	  stars: individual ({\sk}, {\av}) }

\section{Introduction}

The resonance lines of ionized metals are sensitive indicators of the
stellar winds of OB stars, which are rarefied environments
characterized by low degrees of excitation.
Accessible transitions fall in the ultraviolet (UV) and far ultraviolet
(FUV), where they create distinctive P~Cygni profiles by scattering
UV continuum photons.
However, the species available to {\iue} and {\hst} 
-- {\ion{C}{4}}, {\ion{N}{5}}, and {\ion{Si}{4}} --
are not good estimators of the mass-loss rate.
Unsaturated lines are usually from trace ions (e.g., {\ion{Si}{4}}),
so their interpretation depends on the details of the
physics used to model them.
Although {\ion{C}{4}} is a dominant ion over
most of the OB-star domain, carbon is so abundant that the line
is usually saturated and only provides a lower limit on the mass flux.
Consequently, the most reliable mass-loss estimates to date come
from intrinsically weaker wind diagnostics like
H$\alpha$ or free-free emission.

The wealth of resonance lines available in the FUV help to overcome
these problems, since they include diagnostics of abundant
and rare elements that sample a wide range of ionization
potential, including dominant (e.g., {\ion{P}{5}}) and trace
(e.g., {\ion{C}{3}}) ionic species and multiple ionization stages
(e.g., {\ion{S}{4}}/{\ion{S}{6}}; {\ion{P}{4}}/{\ion{P}{5}}).
Previous work with the {\it Copernicus\/} satellite concentrated
on early-type stars in the Galaxy, many of which are severely affected
by interstellar absorption \citep[e.g.,][]{{Snow:1976}, {Walborn:1996}}.
More recent FUV spectra of targets in the Galaxy and the Magellanic Clouds 
were obtained by the Hopkins Ultraviolet Telescope 
\citep[HUT; see, e.g.,][]{Walborn:1995b}
and the Berkeley spectrometer on the {\orfeus} missions
\citep[e.g.,][]{Taresch:1997}.
The launch of the Far Ultraviolet Spectroscopic Explorer
\citep[{\fuse};][]{Moos:2000} revives the possibility of using resonance
lines to determine mass-loss rates for many Galactic and extragalactic
OB stars.

In this Letter we compare {\fuse} spectra of a pair of O7 supergiants,
one in each Magellanic Cloud, and use non-LTE, line-blanketed
model atmospheres to interpret the differences.
The UV spectrum of the target in the LMC, {\sk}, was analyzed
by \citet{Patriarchi:1992}, but otherwise this star has received
little attention.
In contrast, {{\av} = AV~232} is a well-studied object situated
on the edge of the giant {\ion{H}{2}} region NGC~346 in the SMC.
Its UV/optical spectrum has been discussed by \citet{Walborn:1995a},
and a HUT spectrum is described by \citet{Walborn:1995b}.
Basic observational quantities for both targets are listed in
Table~\ref{targets}.

\section{Observations}

Spectra of {\sk} and {\av} were obtained in 1999 September as 
part of the {\fuse} Early-Release Observation program and again
in 1999 October during mirror alignment tests.
These observations were made in time-tag mode through the 
{30\arcsec $\times$ 30\arcsec} (LWRS) aperture \citep[see][]{Moos:2000}
and are characterized by spectral resolution of 12,000--15,000
\citep{Sahnow:2000}.
Since alignment of the four spectrograph channels was not maintained,
the number of channels contributing to the final spectrum and the 
effective exposure times vary with wavelength.
The effective exposure time per channel ranges from 2.7 to 4.9~ksec
for {\sk} and 1.2 to 4.9~ksec for {\av}, and results in measured 
signal-to-noise ratios of 15--20 per resolution element.
The spectra were processed to remove telescope motions during the
alignment tests, collapsed in the spatial dimension, and calibrated
in flux and wavelength.
Model fits to the interstellar H$_2$ spectrum (see \citealt{Shull:2000})
were used to improve the wavelength scale before resampling the
spectra to a constant wavelength step of 0.13~{\AA}.

\begin{figure}
   \epsscale{1.00}
   \plotone{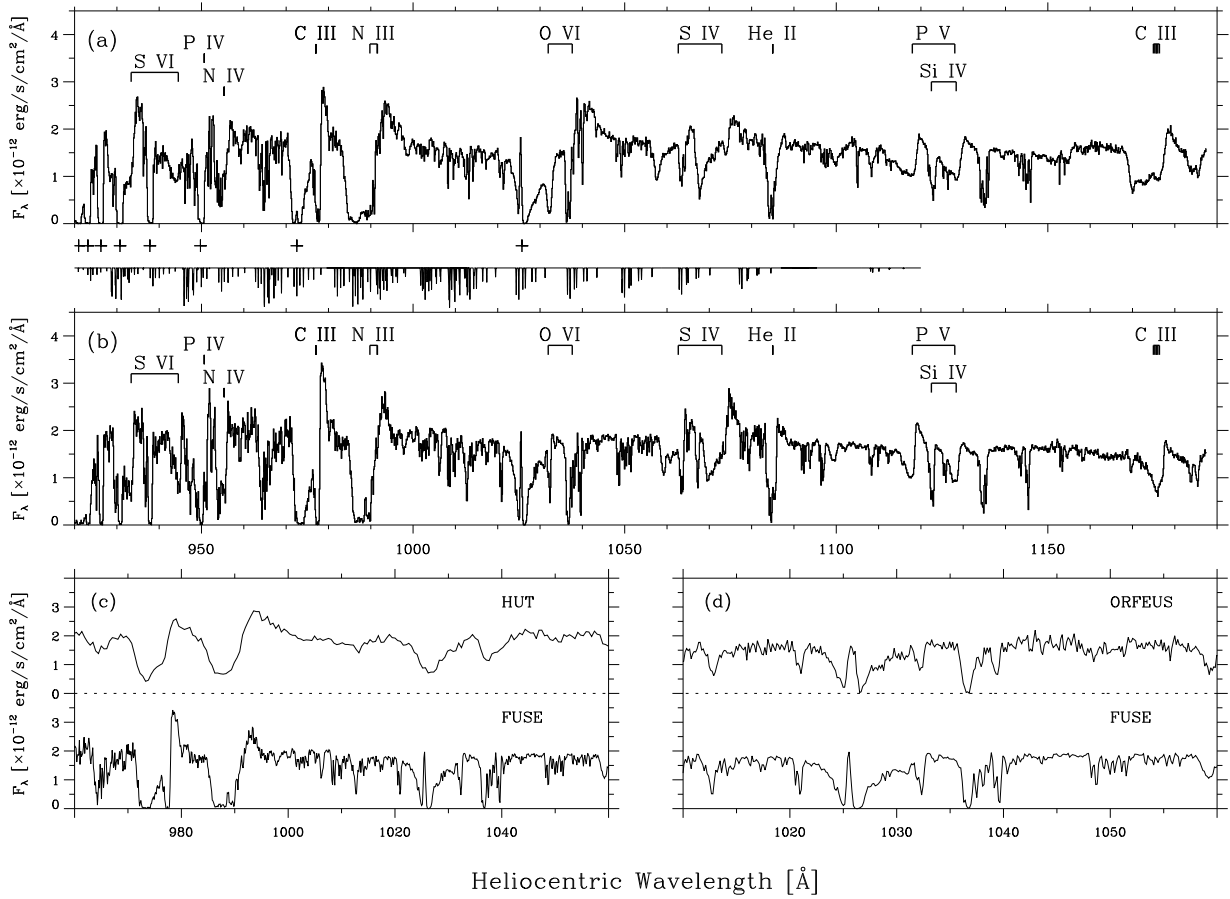}
   \figcaption[Fullerton.Fig1.eps]{
    (a) Calibrated {\fuse} spectrum of {\sk}.
        Rest wavelengths of important stellar transitions are shown.
        The locations and strengths of interstellar H$_2$ lines are
        indicated between panels (a) and (b); crosses mark the
        positions of the Lyman series of {\ion{H}{1}}.
    (b) Same as (a) for {\av}.
    (c) Comparison of spectra of {\av} obtained by HUT and {\fuse}.
    (d) Comparison of spectra of {\av} obtained by {\orfeus} and {\fuse}.
    \label{fusedata}
    } 
\end{figure}

The calibrated spectra are shown in Figure~\ref{fusedata}, along
with identifications of important stellar features and the model
of the interstellar H$_2$ spectrum for {\sk}.
Fig.~\ref{fusedata} also compares the {\fuse} spectrum of {\av} with
spectra obtained by HUT during the Astro-2 mission
\citep{Walborn:1995b} and the Berkeley spectrograph during the
{\it ORFEUS-SPAS II\,} mission \citep{Hurwitz:1998}.
This comparison emphasizes the extremely low background of the {\fuse}
detectors, the reliability of the preliminary flux and wavelength 
calibrations, and the advantages of high spectral resolution for 
disentangling interstellar absorptions from stellar features.

\section{Comparison of Wind Lines}

Two morphological differences between {\sk} and {\av} are 
conspicuous in Fig.~\ref{fusedata}.
First, the {\ion{S}{4}} emission is much stronger in {\av}.
Since {\av} is substantially more luminous than {\sk}, this
difference is likely attributable to the luminosity sensitivity
exhibited by these lines \citep{Walborn:1996}.
Second, {\ion{O}{6}} is barely present in the wind of {\av}: no emission
is visible and only weak absorptions are detectable at the positions
corresponding to the terminal velocity ({\vinf}).
In contrast, the spectrum of {\sk} exhibits a well-formed
{\ion{O}{6}} P~Cygni profile.

Otherwise, the most important difference between the wind lines
in these stars is the smaller {\vinf} exhibited by {\av}; 
see also \citet{Bianchi:2000}.
The theory of line-driven stellar winds predicts that {\vinf}
should decrease with decreasing metallicity \citep{Kudritzki:1989}.
Previous studies with {\iue} and {\hst} have shown that
stars in the SMC follow this trend
(see, e.g., \citealt{{Garmany:1988},{Prinja:1998}}).
After allowing for this systematic difference in breadth,
the absorption and emission strengths of the P~Cygni profiles
are surprisingly similar, despite the substantially lower metal abundances
typical of the SMC.
Some, but not all, of the differences in profile shape are due
to blends with stellar or interstellar features.

\section{Modeling the Spectral Morphology}

As a first step toward understanding the origin of these morphological
differences, we used the program {\sc wm-basic} (v.1.13;
A. W. A. Pauldrach, T. L. Hoffmann, \& M. Lennon, in preparation)
to compute synthetic spectra for both targets.
This program solves the hydrodynamic wind equations for
specified stellar parameters; incorporates non-LTE metal line
blanketing in spherical, expanding geometry; and permits the 
effects of strong shocks in the wind to be incorporated in the
radiative transfer calculations.

We initially adopted {\teff} = 38\,kK for both stars, which is
the mean value for three O7 supergiants (including {\av}) 
determined by \citet{Puls:1996} from non-LTE modeling of photospheric 
lines in the plane parallel, hydrostatic approximation.
Their results for two O7~Ib stars were used to estimate
{\grav} = 3.45 for {\sk}, while their determination of {\grav} = 3.30 
was used for {\av}.
We adopted He/H = 0.10 by number for both stars, and fixed the metal
abundances at 0.4\,$Z_\odot$ and 0.2\,$Z_\odot$ for
{\sk} and {\av}, respectively.
Stellar radii were determined by requiring that the model
fluxes reproduce the dereddened energy distribution of each star.
By using reddening laws for the Galactic foreground \citep{Seaton:1979}
and Magellanic Cloud components \citep{{Howarth:1983},{Bouchet:1985}},
we derived foreground reddening of {\ebv} = 0.04 and 0.03 for
{\sk} and {\av}, respectively, and internal reddenings of
0.05 for both galaxies along these sightlines.
Although the total reddenings are low, they are consistent with the 
\ion{H}{1} column densities of 10$^{21}$cm$^{-2}$ derived for these
stars from fits to the absorption wings of Ly$\alpha$ and $\beta$ 
\citep[e.g.,][]{{Koornneef:1982},{Fitzpatrick:1985}}.

For the adopted {\teff}, {\grav}, and radius, depth independent
wind force multipliers ($\alpha$, $k$, $\delta$) were selected for
{\sc wm-basic} such that the observed values of {\vinf}
(Table~\ref{targets}) were reproduced.
We also required that the mass-loss rate be approximately correct.
For {\av}, we matched the peak intensity of the H$\alpha$
emission profile published by \citet{Puls:1996}.
Since we are not aware of any H$\alpha$ observations of {\sk},
we used the fact that its {\ion{He}{2}}~$\lambda$4686 is
filled in (as implied by its `(f)' classification)
to constrain its mass-loss rate.

\begin{figure}
   \epsscale{1.00}
   \plotone{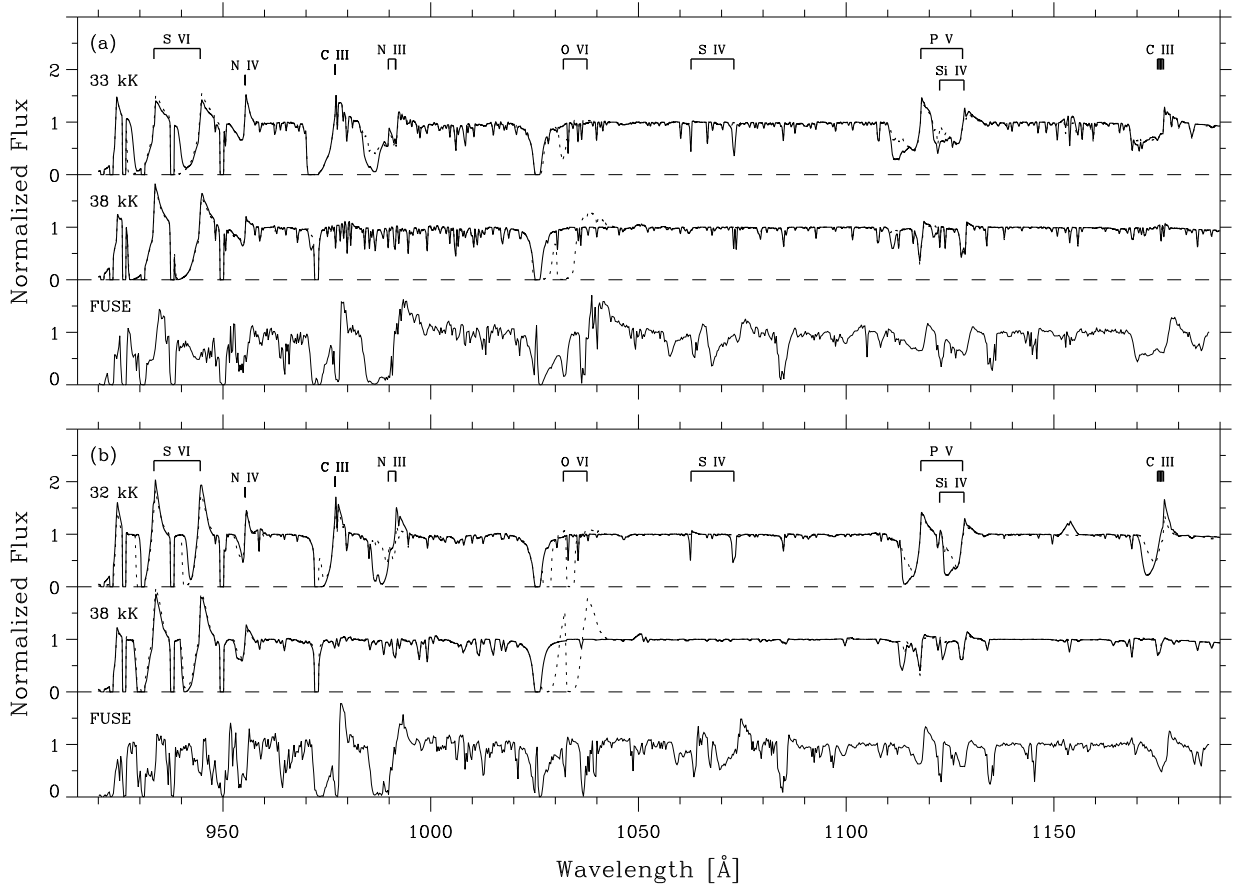}
   \figcaption[Fullerton.Fig2.eps]{
   (a) Comparison between {\fuse} observations and synthetic spectra for 
       {\sk}.
       The computed spectra are labelled by the {\teff} of the model;
       dashed lines indicate the spectrum when the effects of shocks
       are included.
       Absorption by a column density of 10$^{21}$\,cm$^{-2}$ of \ion{H}{1}
       has been included in the models.
   (b) Same as (a) for {\av}.
    \label{fusemodels}
    }
\end{figure}

Figure~\ref{fusemodels} compares the {\fuse} spectra of the
targets with synthetic spectra computed for {\teff} = 38\,kK.
The models do not reproduce the dominant wind features in the FUV 
spectra of these stars.
In particular, they do not predict the occurrence of
{\ion{C}{3}}, {\ion{N}{3}}, or {\ion{S}{4}} to any significant
degree.
The absence of these ions implies that the initial wind models are too hot.
We therefore computed cooler models characterized
by {\teff} = 33\,kK and 32\,kK for {\sk} and {\av}, respectively.
The synthetic spectra from these models are also shown in
Fig.~\ref{fusemodels}.
As expected, the cooler temperatures improve the correspondence between
the appearance of the {\ion{C}{3}} and {\ion{N}{3}} lines, and produce
only minor changes in the {\ion{S}{6}} resonance lines.
Although the agreement is much better, the cooler models predict {\ion{P}{5}}
profiles that are too strong and {\ion{S}{4}} lines that are much
weaker than observed.

Neither model predicts the occurrence of {\ion{O}{6}} in the wind of {\sk}.
To explore possible mechanisms for the production of this ion,
we incorporated the extreme-UV radiation field due to strong shocks,
which could be caused by the line-driven instability or the
presence of other time-dependent structures in the wind 
\citep[see, e.g.,][]{Owocki:1988}.
We used the characterization of shock phenomena implemented in
{\sc wm-basic} with assumed input parameters of
$L_{\rm X}/L_{\rm bol} \sim 10^{-7}$ \citep{Chlebowski:1989} and
$v_{\rm turb}$/{\vinf} = 0.12.
Synthetic spectra from these models are included in Fig.~\ref{fusemodels}
as dotted lines.
As expected, the presence of shocks increases the strength of the
{\ion{O}{6}} lines in all models.
At the cooler {\teff}, the shocks also decrease the strength
of {\ion{N}{3}} and {\ion{P}{5}} substantially.
The strength of the {\ion{O}{6}} emission lobe of {\sk} is not reproduced
at {\teff} = 33\,kK.

Although the modeling is still preliminary, it is clear that
models with {\teff} $\sim$32\,kK are better matches to the {\fuse}
spectra of {\sk} and {\av} than models with {\teff} $\sim$38\,kK.
Provisional values for the parameters of the cooler models
are listed in Table~\ref{targets}.
Even though O stars in the SMC are predicted to have weaker
winds than their counterparts in the LMC, the models suggest that
the mass-loss rate of {\av} exceeds the rate of {\sk} by
a factor of $\sim$3.
Evidently the greater luminosity of {\av} more than offsets any
reduction in radiative driving due to reduced metallicity.
Although we have not yet computed model grids for different
abundances of C and N that might arise from rotational
mixing of CNO-cycled material into the atmosphere \citep{Venn:1999}, 
test calculations indicated that improved fits 
to the {\ion{N}{3}} resonance line produce worse matches in the {\ion{C}{3}}
lines, since N enrichment is accompanied by C depletion.

\section{Discussion}

The {\fuse} observations of {\sk} and {\av} represent both a challenge
to state-of-the-art model atmospheres and a unique opportunity to probe
the physics of hot-star winds.
We have identified two specific problems.

First, initial analysis of the FUV spectra suggests that the temperatures
of both stars are lower than the values derived from fits to photospheric
lines by {$\Delta T \approx -5$\,kK}.
\citet{Crowther:1997} also found much cooler temperatures (by $\sim$8\,kK)
from their analysis of the optical wind lines of two Galactic O8
supergiants.
Thus, the currently adopted temperatures for late O supergiants
may be systematically too high.
Since \citet{Herrero:2000} found that fits to Galactic O6 stars made with
unified model atmospheres require lower temperatures than previous 
analyses (by 3 and 4.5\,kK for a dwarf and supergiant, respectively),
at least part of this discrepancy may be attributed to the 
inadequacies of plane-parallel model atmospheres.

Second, even line-blanketed wind models with {\teff} = 38\,kK
do not predict that much {\ion{O}{6}} is present in the spectra of
O7 supergiants.
Instead, the strong {\ion{O}{6}} P~Cygni profile exhibited by {\sk}
must be due to the presence of shocks from time-dependent structures
in the wind.
\citet{Pauldrach:1994} and \citet{Taresch:1997} reached the same
conclusion from their analyses of the FUV and UV spectra of hotter 
O supergiants.
The absence of a comparable {\ion{O}{6}} feature in {\av} is
puzzling: is it a direct consequence of the reduced metal abundance,
or are the shocks intrinsically weaker in {\av} because the wind
velocities are smaller?

{\fuse} will observe a large sample of OB stars in the Milky Way,
LMC, and SMC in order to address these and other issues.
A variety of complementary analysis techniques will be used to 
interpret these data, which will be combined with archival
{\iue}, {\hst}, and optical data whenever possible.
We anticipate that this work will return qualitatively new information
about the properties of line-driven stellar winds.

\acknowledgments
This work is based on data obtained for the Guaranteed Time Team by
the NASA-CNES-CSA {\fuse} mission operated by the Johns Hopkins University.  
Financial support from NASA contract NAS5-32985 
(U. S. participants), the CSA Space Science Program (AWF), the
Royal Society (PAC), and PPARC (OD) is acknowledged.  
We are grateful to Dr. Adi Pauldrach for the use of {\sc wm-basic},
Dr. Van Dixon for providing the {\orfeus} spectrum of {\av},
and Roger Wesson for deriving initial stellar parameters.


\clearpage
\begin{deluxetable}{lccccccccccc}

 \tabletypesize{\scriptsize}
 \tablecolumns{12}
 \tablewidth{0pc}
 \tablecaption{Target Stars\label{targets}}
 \tablehead{
 \colhead{Object}                        &           
 \colhead{Galaxy}                        &           
 \colhead{Sp. Type}                      &           
 \colhead{$V$}                           &           
 \colhead{\ebv}                          &           
 \colhead{$M_{\rm V}$\tablenotemark{a}}  &           
 \colhead{\vinf}                         &           
 \colhead{{\teff}}                       &           
 \colhead{$R_{\ast}$}                    &           
 \colhead{\grav}                         &           
 \colhead{$\log L/L_{\odot}$}            &           
 \colhead{$\log \dot{M}$}                \\          
                                         &           
					 &           
                                         &           
                                         &           
                                         &           
                                         &           
 \colhead{[\kms]}                        &           
 \colhead{[kK]}                          &           
 \colhead{[$R_{\odot}$]}                 &           
                                         &           
                                         &           
 \colhead{[\msunpyr]}                    \\          
}
 \startdata
 \sk                                     &           
 LMC                                     &           
 O7~Ib(f) (1)                            &           
 12.57                                   &           
 0.09 (3)                                &           
 $-$6.3                                  &           
 1800$\pm$100 (4)                        &           
 33                                      &           
 22                                      &           
 3.3                                     &           
 5.7                                     &           
 $-$5.7                                  \\[5pt]     
 \av                                     &           
 SMC                                     &           
 O7~Iaf$+$ (2)                           &           
 12.36                                   &           
  0.08 (3)                               &           
 $-$6.8                                  &           
 1400$\pm$100 (4)                        &           
 32                                      &           
 29                                      &           
 3.1                                     &           
 5.9                                     &           
 $-$5.3                                  \\          
                                         &           
                                         &           
                                         &           
                                         &           
                                         &           
\multicolumn{2}{l}{\citealt{Puls:1996}:} &           
 37.5                                    &           
 29.3                                    &           
 3.3                                     &           
 6.2                                     &           
 $-$5.3                                  \\          
 \enddata
 \tablerefs{ (1) \citealt{Fitzpatrick:1988};
             (2) \citealt{Walborn:1977};
             (3) this work;
             (4) \citealt{Bianchi:2000}.
           }
 \tablenotetext{a}{Derived assuming distance moduli of 18.5 and 18.9
                   for the LMC and SMC, respectively 
                   \citep{Westerlund:1997}.}
\end{deluxetable}

\clearpage

\clearpage

\end{document}